\documentstyle[LTpaper]{article}
\begin{document}
\input{epsf}
\title{Theory of short-range magnetic order for the t-J model}
\author{C. Schindelin$^a$, U. Trapper$^b$, H. Fehske$^a$, and H. B\"uttner$^a$}
\address{\mbox{$^a$ Physikalisches Institut, Universit\"at Bayreuth, 
D-95440 Bayreuth, Germany}
\vskip 10pt \mbox{$^b$  
Institut f\"ur Theoretische Physik, Universit\"at Leipzig, 
D-04109 Leipzig, Germany}}
\abstract{
We present a self-consistent theory of magnetic short-range order  
based on a spin-rotation-invariant slave-boson  representation of the 
2D t-J model. In the functional-integral scheme, 
at the nearest-neighbour pair-approximation level, 
the bosonized t-J Lagrangian is transformed to a classical
Heisenberg model with an effective (doping-dependent) 
exchange interaction which takes into account the interrelation of
``itinerant'' and ``localized'' magnetic behaviour. Evaluating the theory 
in the saddle-point approximation, we find a suppression of
antiferromagnetic and incommensurate spiral long-range-ordered phases
in the favour of a paramagnetic phase with pronounced
antiferromagnetic short-range correlations.
}
\maketitle\pagestyle{empty}
Experimental evidence has been accumulating that high-$T_c$
superconductivity in the perovskite copper oxides develops in the
presence of strong antiferromagnetic (AFM) spin
correlations, which may have important implications
for the pairing mechanism. The interesting low-temperature
magnetic behaviour comes predominantly from the complicated 
interplay between {\it itinerant} charge carriers (holes) and
{\it localized} spins $(\rm Cu^{2+})$ within the $\rm CuO_2$
planes. Hole doping rapidly destroys the AFM long-range order
(LRO), but pronounced {\it short-range order} (SRO) is retained and may
account for the unusual normal-state properties of the cuprates, 
such as the behaviour of the uniform magnetic susceptibility 
as a function of doping and temperature~\cite{Jo89}.
 
Motivated by this situation, in this article we   
outline a theory of magnetic SRO 
for strongly correlated electron
systems  described by the 2D t-J model
\begin{equation}
{\cal H}=-t \sum_{\langle i j \rangle \sigma} 
\Big(\tilde{c}_{i\sigma}^\dagger 
\tilde{c}_{j\sigma}^{} + {\rm H.c.}\Big)+ J \sum_{\langle i j\rangle}
\Big(\vec{S}_i^{}\vec{S}_j^{} - \frac{1}{4}\tilde{n}_i^{}\tilde{n}_j^{}\Big)\,.
\end{equation}
Applying a spin-rotation-invariant slave-boson (SB) technique 
within the functional-integral representation of the 
partition function~\cite{DF94}, the bosonized free--energy
functional of the t-J model takes the form
\begin{eqnarray}
{\mit \Psi} &=& \sum_i (-\nu_i n_i + \vec{\xi}_i \vec{m}_i)  
- \frac{1}{\beta} \mbox{Tr}_{ij,n\atop \rho \rho' \;\;} 
\ln[-\hat{G}^{-1}] \nonumber\\ 
& & \qquad + \frac{J}{4} \sum_{\langle ij\rangle}(\vec{m}_i \vec{m}_j 
 - n_i n_j )
\end{eqnarray}
with the transformed inverse propagator (cf. Ref.~\cite{TIF95})
\begin{eqnarray}
\underline{\hat{G}}_{ijn}^{-1}
= (\underline{z}_i 
\underline{z}_j)^{-1} [ (-i\omega_n - \nu_i) \underline{1} + 
\vec{\xi}_i \underline{\vec{\sigma}} ] \delta_{ij} - t_{ij}
\underline{1}\,.
\end{eqnarray}
(underbars denote a $2\times 2$ matrix in spin space). 
Here, the local magnetization [particle number]
operators are given by $\vec{m}_i = 2\,p_{io}\,\vec{p}_i$ 
$[n_i = p_{io}^2 + \vec{p}_i^{\,2}]$, $\;\vec{\xi}_i$ 
$[\nu_i]$ refer to  the ``internal'' magnetic [charge] fields,
and the nonlinear $(\underline{z}_i)$--factors, 
which depend only on the single-occupancy matrix operators 
\mbox{$\underline{p}_i \!=\! \frac{1}{2}\!(\underline{1} 
p_{io} \!+\! \underline{\vec{\sigma}}  \vec{p_i} 
)$}, yield a correlation-induced band renormalization~\cite{DF94}.
For simplicity, we describe the fluctuations of the bosonic fields by
the Ansatz ($|\vec{s}_i|=1$):
\begin{eqnarray}
n_i = n\quad\; & &\vec{m}_i = \bar{m}\, \vec{s}_i \\
\nu_i = \nu\quad\; & & \vec{\xi}_i = \bar{\xi} \, \vec{s}_i\,,
\end{eqnarray}
i.e., we assume that the charge fields as well as the amplitudes of the spin
components are site independent. Moreover, since the flipping time 
of the local magnetizations is supposed to be long compared to the
electronic hopping time all bosonic degrees of freedom are treated 
in the static approximation. 

To incorporate SRO effects, one has to go beyond the homogeneous 
paramagnetic (PM) saddle--point. Therefore we perform an 
expansion in terms of the local perturbation
%
$\underline{V}_i \delta_{ij}\!\! =\!\!-\underline{\hat{G}}_{ij}^{-1}
\!\!+ \underline{\hat{G}}_{ij}^{o\: -1}$,
\mbox{where the} PM propagator $\underline{\hat{G}}_{ij}^{o\: -1}$, 
with the diagonal [off--diagonal] components
$ \hat{G}_{0}^{o}\equiv\hat{G}_{ii}^{o}$ 
[$\hat{G}_{1}^{o}\equiv \hat{G}_{\langle ij\rangle}^{o}$],
arises out of (3) by setting $\vec{\xi}_i=0$ and 
replacing $\underline{z}_i\!\to\! z^o$, $\nu_i\!\to\!\nu^o$.
Using spherical harmonics we are able to transform  
${\mit \Psi}(\{\vec{s}_i\})$~\cite{TIF95} 
to an {\it effective} classical Heisenberg model 
(within the nearest--neighbour pair approximation): 
\begin{equation}
{\mit \Psi} = \bar{{\mit \Psi}} - \bar{J} 
\sum_{\langle ij\rangle } \vec{s}_i \vec{s}_j\,,
\end{equation} 
where
\begin{eqnarray}
\bar{{\mit \Psi}} / N\!  \!&=&\!  \!{\mit \Psi}^o/N 
-\nu \, n + \bar{\xi} \, \bar{m} - \frac{J n^2}{2} 
\nonumber\\ & &
+ \sum_{\zeta} \left[ {\mit \Phi}_{0\zeta} 
+  \int_{-1}^1 dx \: {\mit  \Phi}_{1\zeta}(x) \right]\,,\\
\bar{J} \!\!  &=&\!\!  - \frac{J \bar{m}^2}{4} 
- \frac{3}{2} \sum_{\zeta} \int_{-1}^1 dx\:
x\: {\mit \Phi}_{1\zeta}(x)\,, \\
{\mit \Psi}^o\!\! &=&\!\!  - \frac{2}{\beta} \sum_{\vec{k}} \ln\,
[ 1 + e^{-\beta \{(z^0)^2 \varepsilon_{\vec{k}} +\nu^o - \mu\}}],
\\
{\mit \Phi}_{0\zeta}\!\!  &=&\!\!  \mbox{Tr}_{n} \ln \,
[1 -\hat{G}_0^o V_{\zeta}]\,,
\\  
{\mit \Phi}_{1\zeta}\!\!  &=&\!\!  \mbox{Tr}_{n} \ln \,
[1 - (\hat{G}_1^o)^2 T^{(2)}_{\zeta} (x)]\,,
\end{eqnarray}
$x=\vec{s}_0 \vec{s}_1$, $\underline{T}_i = 
\underline{V}_i (\underline{1} - \hat{G}_0^o \underline{V}_i)^{-1} $, 
and $V_{\zeta}$ and $T_{\zeta}^{(2)}(x)$
are the eigenvalues of $V_{i\rho\rho'}$ and
$\left( \underline{T}(\vec{s}_0) 
\underline{T}(\vec{s}_1) \right)_{\rho \rho'}$, respectively.
Note that the effective Heisenberg--exchange integral $\bar{J}$   
has to be determined self--consistently at each given interaction
strength $J$ and hole doping $\delta=1-n$.  Evaluating the
trace over the $\vec{s}_i$-variables in the partition function~\cite{Cu95} 
and hereafter 
adopting the saddle-point approximation to the resulting {\it nonlocal} 
Heisenberg free--energy functional,
\begin{equation}
{\mit \Psi} = \bar{\mit \Psi} - \frac{2N}{\beta} \ln \,
\left[ 4 \pi \frac{\sinh(\beta \bar{J})}{\beta \bar{J}}\right]\,,
\end{equation} 
the extremal Bose fields are obtained from 
\begin{eqnarray}
n = \sum_{\zeta} n_{\zeta}^f
& &\!\!\!\!
\nu = \sum_{\zeta} b_{\zeta}^f \frac{\partial [z^2]_{\zeta}}{\partial n} - 
Jn
\\ 
\!\!\!\!\!\bar{m} = \sum_{\zeta} \zeta n_{\zeta}^f
& &\!\!\!\!
\bar{\xi} = - \sum_{\zeta} b_{\zeta}^f 
\frac{\partial [z^2]_{\zeta}}{\partial \bar{m}} -
\eta \frac{J}{\bar{m}}  
\end{eqnarray}
with $x_{\zeta}^f = \frac{\partial \phi_{0\zeta}}
{\partial (y_\zeta)} \!+\!\sum_{\zeta^{\prime}}\int\!dx
\left[ 1 \!+\!\eta\frac{3x}{\bar{m}^2} \right]\!
\frac{\partial \phi_{1\zeta^{\prime}}(x)}{\partial (y_\zeta)}$, where
$y_\zeta\equiv(\nu-\zeta\bar{\xi})\;$ $[y_\zeta\equiv(z^2)_{\zeta}]$ 
if $x\leftrightarrow n\,$ [$x\leftrightarrow b$], and 
we have introduced the SRO parameter
\begin{equation}
\eta\equiv\langle \vec{m}_i \vec{m}_j  \rangle =\bar{m}^2
\mbox{L}(\beta\bar{J}) 
\end{equation} 
with the Langevin function  
$\mbox{L}(z)= \coth z
-z^{-1}$.
For vanishing $\bar{m}$, we have $\bar{J}=0$, and the PM saddle-point
is recovered, i.e., our theory adequately describes
paraphases without and with SRO $(\vec{m}=\langle\vec{m}_i\rangle)$:
\begin{equation}
\begin{array}{c@{\;:\;\;\;\;\;} l@{\;,\;\;\;\;\;} l}
\mbox{PM} &\vec{m}=0& \eta = 0  \\
\mbox{SRO-PM}& \vec{m}=0 & \eta \neq 0\;.  
\end{array}
\end{equation}
\begin{figure}[t]
\begin{minipage}[t]{8cm}
\centerline{\mbox{\epsfxsize 7cm\epsffile{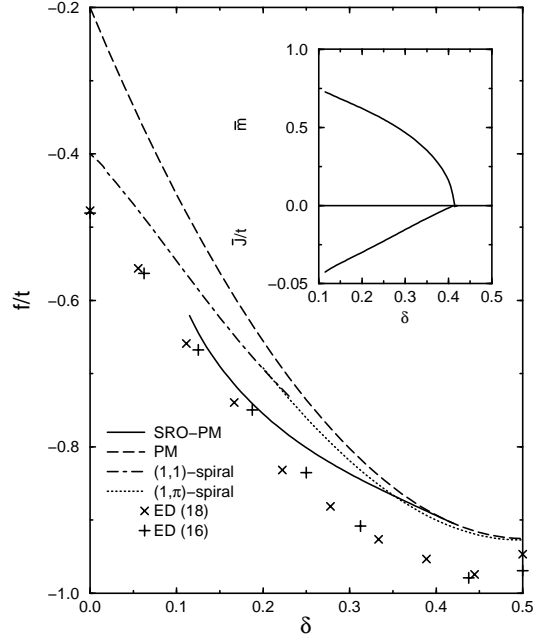}}}
\caption{Ground-state energies of the 2D t-J model as functions of
doping at $J/t=0.4$. The energy of the  SRO-PM phase is 
compared with SB results for the PM and different spiral states as well as
with exact diagonalization (ED) data obtained for the 16- and 18-site
lattices. The corresponding local magnetization $\bar{m}$ and  
effective AFM exchange coupling  $\bar{J}$ are shown in the inset.}
\label{fig1}
\end{minipage}
\end{figure}
To illustrate the quality of our approach, some representative numerical
results are depicted in Fig.~1. Obviously, in the physically 
most interesting doping region 
the ground-state energy of the SRO-PM phase is
lower than that of the frequently discussed 
spiral phases and lies close to the exact data. Thus  
upon doping magnetic LRO make way to SRO. 
Note that the SRO-PM phase is locally stable against phase
separation. The interplay of local and itinerant
magnetic behaviour, which is self-consistently incorporated in our
theory, results in strong doping (and temperature) 
dependences of both $\bar{m}$ and $\bar{J}$.


\begin{thebibliography}{99}
\bibitem{Jo89}
\mbox{J. $\!$Torrance~et $\!$al.,$\!$ Phys.$\!$ Rev.$\!$ B$\!$ {\bf
40},$\!$ (1989)$\!$ 8872.}
\bibitem{DF94}
M. Deeg and H. Fehske, Phys. Rev. B {\bf 50} (1994) 17874.
\bibitem{TIF95}
U. Trapper, D. Ihle, and H. Fehske, Phys. Rev. B {\bf 52} (1995) R
11553.
\bibitem{Cu95}
J. Cur\'{e}ly, Europhys. Lett. {\bf 32} (1995) 529.
\end{thebibliography}
\end{document}